\documentclass[]{aastex631}

\newcommand{\Msun}{M_\odot}

\begin{document}

\title{Quantifying the evidence against a mass gap between black holes and neutron stars}

\correspondingauthor{J.E. Horvath}
\email{foton@iag.usp.br}

\author[0000-0003-3109-9042]{L.M. de Sá}
\affiliation{Universidade de São Paulo (USP), Instituto de Astronomia, Geofísica e Ciências Atmosféricas (IAG), \\
R. do Mat\~ao 1226 - Cidade Universit\'aria,
05508-090, S\~ao Paulo - SP, Brazil}

\author[0000-0002-5914-0556]{A. Bernardo}
\affiliation{Universidade de São Paulo (USP), Instituto de Astronomia, Geofísica e Ciências Atmosféricas (IAG), \\
R. do Mat\~ao 1226 - Cidade Universit\'aria,
05508-090, S\~ao Paulo - SP, Brazil}

\author[0000-0002-1420-8991]{R.R.A. Bachega}
\affiliation{Universidade de São Paulo (USP), Instituto de Astronomia, Geofísica e Ciências Atmosféricas (IAG), \\
R. do Mat\~ao 1226 - Cidade Universit\'aria,
05508-090, S\~ao Paulo - SP, Brazil}

\author[0000-0003-4089-3440]{J.E. Horvath}
\affiliation{Universidade de São Paulo (USP), Instituto de Astronomia, Geofísica e Ciências Atmosféricas (IAG), \\
R. do Mat\~ao 1226 - Cidade Universit\'aria,
05508-090, S\~ao Paulo - SP, Brazil}

\author[0000-0003-4543-0912]{L.S. Rocha}
\affiliation{Universidade de São Paulo (USP), Instituto de Astronomia, Geofísica e Ciências Atmosféricas (IAG), \\
R. do Mat\~ao 1226 - Cidade Universit\'aria,
05508-090, S\~ao Paulo - SP, Brazil}
\affiliation{Max-Planck-Institut für Radioastronomie, \\ Auf dem Hügel 69, 53121 Bonn, Germany}

\author[0000-0002-8478-5460]{P.H.R.S. Moraes}
\affiliation{Universidade Federal do ABC (UFABC), Centro de Ciências Naturais e Humanas (CCNH), \\
Avenida dos Estados, 5001, 
09210-580, Santo Andr\'e - SP, Brazil}

\begin{abstract}

The lack of objects between $2$ and $5 \, M_{\odot}$ in the joint mass distribution of compact objects has been termed  ``mass gap'', and attributed mainly to the characteristics of the supernova mechanism precluding their birth. However, recent observations show that a number of candidates reported to lie inside the ``gap'' may fill it, suggesting instead a paucity that may be real or largely a result of small number statistics. We quantify in this work the individual candidates and evaluate the joint probability of a mass gap. Our results show that an absolute mass gap is not present, to a very high confidence level. It remains to be seen if a relative paucity of objects stands in the future, and how this population can be related to the formation processes, which may include neutron star mergers, collapse of a neutron star to a black hole and others.

\end{abstract}


\section{Introduction} \label{sec:intro}

The mass distribution of compact objects has long been a topic of central importance to the fields of late stellar evolution and dense matter physics, with implications for the mechanisms of supernova explosions and the internal structure of neutron stars (NSs), as well as their maximum mass, $M_\mathrm{max}^\mathrm{NS}$. One of the important issues within this topic has been that of the putative existence of a {\it mass gap} between the heaviest NSs at $\sim M_\mathrm{max}^\mathrm{NS}$, and the lightest observed black holes (BHs), at $\sim 5\,\Msun$. The idea originally arose in \citet{Bailyn1998}, who performed a Bayesian analysis of a sample of 7 BH-hosting candidate low-mass X-ray binaries and determined the existence of a gap in the distribution between the extreme upper limit of $3\,\Msun$ for NS masses \citep[essentially the limit by][]{Rhoades1974} and the least massive BH in their sample. Further studies of the BH mass distribution continued to support the existence of a gap by finding a minimum BH mass substantially larger than the maximum NS mass \citep{Ozel2010, Farr2011}, and the presence of this feature has been linked to the convection growth timescales in supernovae \citep{Fryer_2012,Belczynski_2012}. The idea of an absence of objects inside the gap will be called a {\it desert} (or absolute) gap in the rest of this work.

Work has been done thereafter to constrain the bounds of this mass gap \citep[called ``lower'' to distinguish it from the ``upper'' observed in the massive range of BH distribution,][]{Woosley2021}. On one side, this has consisted in studying the NS mass distribution and $M_\mathrm{max}^\mathrm{NS}$, tightly related to the gap lower limit, $M_\mathrm{lower}$, which have yielded results for $M_\mathrm{max}^\mathrm{NS}$ in the $2.2-2.6\,\Msun$ range \citep{Margalit_2017,Ruiz2018,Alsing2018,Rezzolla2018,Shibata2019,Ai_2020,Shao2020,Rocha2021}. On the other hand, effort has been spent on reconstructing the stellar-mass BH distribution with more recent data and looking for evidence of a minimum stellar BH mass, yielding values between $\sim 4-5\,\Msun$ for samples of X-ray binaries \citep{Ozel2010,Farr2011} and between $\sim 4-6\,\Msun$ from gravitational wave data \citep{Fishbach2020, Abbott2021a, Farah2022}.

It has also been claimed that a scarcity of ``mass-gap objects'' may be just due to observational limitations and systematic errors, instead of a physical feature of compact object formation mechanisms. For example, \cite{Kreidberg2018} have shown that a common assumption of zero or constant emission from accretion flow in X-ray transients can lead to systematic underestimation of the orbital inclination, and that correcting for this leads to lower masses, reaching the gap range. Gravitational observations might be limited as well when addressing the gap region, given the lower detector sensitivities to the quieter low-mass events. According to this alternative picture, if a physical gap in the mass distribution does exist, it may be found for specific classes of systems only, such as X-ray binaries, but not for the overall compact object population; it could be expected then that as our capacity of observing compact objects with different origins grows, the gap in the full distribution will be filled to some extent by particular formation channels, while remaining a feature of others.

Recent detections and mass measurements of compact objects in a variety of systems have suggested that the $\sim2-5\,\Msun$ interval might indeed be at least partially filled, which would indicate a {\it depleted} (relative) instead of ``desert'' (absolute) gap. These include gravitational wave detections of compact binary mergers, in which either the progenitors or the remnant were within the gap range \citep{Shibata2019, Abbott2020a, Abbott2021a, GWTC21, GWTC3}; observations of detached binaries \citep{Thompson2019, Jayasinghe2021}; of spider binaries \citep{Linares2018, Romani2022}; and microlensing events \citep{Lam2022}. In this work, in order to test the hypothesis of a depleted gap against a desert one, we evaluate the impact of these new observations against current NS and BH joint mass distributions that feature a desert gap. Because the minimum mass in which a gap could start is still uncertain, we allow for variation in the lower limit of the gap and study the probability that the absolute gap is real as a function of its lower limit. Then, we investigate whether the recent observations are compatible with the desert/absolute gap model from a statistical standpoint.

We describe the mass distribution models employed in Section \ref{sec:mass_distr}. In Section \ref{sec:gap_objects} we list the recent gap object candidates and discuss the treatment of the uncertainties for the mass of each of them. In Section \ref{sec:methods} we describe the statistical methods used to confront the depleted and desert gap hypotheses, and in Section \ref{sec:results} show that the new objects are incompatible with the desert gap model, excluding it with a high degree of confidence. Our concluding remarks are presented in Section \ref{sec:conclusion}. In Appendix \ref{app:additional_gap} we list the mass measurements of an extended set of gap candidates, including weaker candidates and less precise mass estimates. In Appendix \ref{app:bhnsdistr} we list the BH and NS masses used in determining the starting compact object mass distributions.

\section{Mass distribution}
\label{sec:mass_distr}

While supernova engine prescriptions \citep[and references therein]{Patton2022} and recent studies of the compact object mass distribution \citep{Fishbach2020, Farah2022} have treated it as a single, continuous distribution, spanning both the NS and BH ranges, we consider first the distributions for each class of objects separately, allowing for some overlap in the gap region. In fact, some works have shown a non-monotonic formation of NSs and BHs \citep{Sukh2016}, suggesting that the inference of a joint distribution from simulation work is complex and dependent on the input physics. 

For the NS distribution ($\mathcal{D}_\mathbf{NS}$) we employ the double Gaussian from \cite{Rocha2021}, given in Table II therein, for a sample of 95 measured NS masses \footnote{At the time of submission, \cite{Rocha2021} relate 96 masses. However, one of the objects is duplicated in their sample. See Appendix \ref{app:bhnsdistr} for more.}. This consists of the large amplitude peak modeled by a Gaussian with $\mu_1=1.365\,\Msun$, $\sigma=0.109\,\Msun$ and weight $r_1=0.498$, that is, a frequent value found in NSs; and a second lower amplitude peak tentatively associated to born massive and significantly accreted NSs, modeled by a Gaussian with $\mu_2=1.787\,\Msun$ and $\sigma_2=0.314\,\Msun$. Because we do not discriminate between NSs and BHs among the gap candidates, we have chosen a non-truncated distribution which in principle allows for the possibility of gap NSs, although \cite{Rocha2021} show that only $\sim 7\%$ of these would have mass greater than $2.59\,\Msun$, their $M_\mathrm{max}^\mathrm{NS}$ found for the truncated model (Table I therein).

For the BH distribution ($\mathcal{D}_\mathbf{BH}$) we adopt the simple Gaussian from \citet{Ozel2010}, which found that their sample of $16$ low-mass X-ray binaries is consistent with a narrow distribution with $\mu=7.8\,\Msun$ and $\sigma = 1.2\,\Msun$. This model does not incorporate recent BH mass measurements, such as the rich catalog built up from gravitational-wave observations \citep{GWTC3}, nor the possible systematic errors in BH mass estimates discussed by \cite{Kreidberg2018}. Both will be taken into account for an updated revised distribution (Bernardo et al., in preparation). For the present work, we expect our analysis not to be strongly affected by the recent observations as these have predominantly populated the $\geq10\,\Msun$ range. Particular observations that have resulted in masses in or near the gap region are included as gap candidates in Section \ref{sec:gap_objects}, or mentioned in Appendix \ref{app:additional_gap} in the case of unconfirmed or weaker candidates.

\begin{figure}[ht!]
\plotone{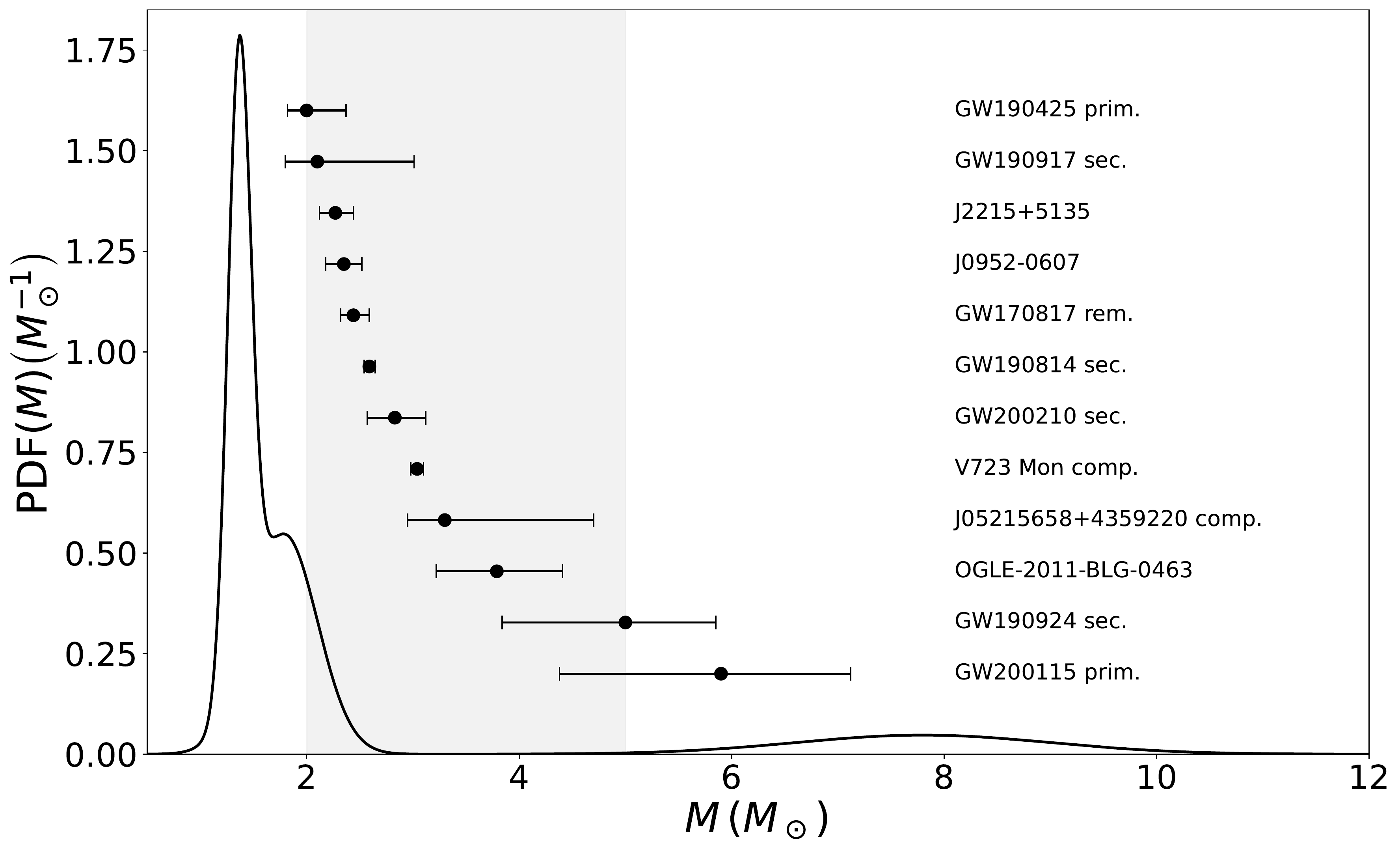}
\caption{Joint empirical distributions for compact object masses as the Gaussian mixtures of the fits for NSs from \cite{Rocha2021} and BHs from \cite{Ozel2010}, weighted by sample sizes (95 NSs and 16 BHs), with the putative gap region between $2$ and $5\mathrm{M}_\odot$ highlighted in a darker shade. Also shown in the gap between the BH and second NS peaks are the nominal masses and $1\sigma$ confidence intervals for the candidate gap objects from Table \ref{tab:newobjects}.
\label{fig:distributions_and_objects}}
\end{figure}

We take as our fiducial, desert gap, distribution ($\mathcal{D}_\mathrm{desert}$) the Gaussian mixture of $\mathcal{D}_\mathrm{NS}$ and $\mathcal{D}_\mathrm{BH}$, weighted by their respective sample sizes, which is shown in Figure \ref{fig:distributions_and_objects} together with the individual gap candidates. We do not take into account uncertainties in the distribution parameters. The mass distributions for the candidates are discussed in Section \ref{sec:gap_objects}. A full account of the measurements considered in fitting the NS and BH mass distributions is provided in Appendix \ref{app:bhnsdistr}.

\section{Candidate gap objects}
\label{sec:gap_objects}

We list in Table \ref{tab:newobjects} the 12 mass gap object candidates considered in this work, ordered by their nominal masses, which also appear in Figure \ref{fig:distributions_and_objects}. This sample consists of recent mass measurements from companion light curves, binary mergers and one microlensing event. Their individual mass distributions are treated either as Gaussians or asymmetrical Gaussians according to each particular case. We have chosen not to include a subset of eight additional candidates, which are listed in Appendix \ref{app:additional_gap}.

\begin{deluxetable*}{lcl}[h]
\tabletypesize{\scriptsize}
\tablewidth{0pt} 
\tablecaption{Mass Distributions for the 12 New Objects Used in This Work\label{tab:newobjects}}
\tablehead{
\colhead{Name} & \colhead{$M$}& \colhead{Reference} \\
{} & $\left(\Msun\right)$ & {}
} 
\startdata 
GW 200115 prim. & $AN(5.9, 1.22, 1.52)$ & \cite{GWTC3} \\
GW 190924 sec. & $AN(5, 0.85, 1.16)$ & \cite{GWTC2} \\
OGLE-2011-BLG-0463 (DW) & $AN(3.79, 0.62, 0.57)$ & \cite{Lam2022} \\
OGLE-2011-BLG-0463 (EW) & $AN(2.15, 0.67, 0.54)$ & \cite{Lam2022} \\
2MASS J05215658+4359220 comp. & $AN(3.3, 1.4, 0.35)$ & \cite{Thompson2019} \\
V723 Mon comp. & $N(3.04, 0.06)$ & \cite{Jayasinghe2021} \\
GW 200210 sec. & $AN(2.83, 0.29, 0.26)$ & \cite{GWTC3} \\
GW 190814 sec. & $N(2.59, 0.05)$ & \cite{Abbott2020a} \\
GW 170817 rem. & $AN(2.44, 0.15, 0.12)$ & \cite{Shibata2019}\\
PSR J0952-0607 & $N(2.35, 0.17)$ & \cite{Romani2022} \\
PSR J2215+5135 & $AN(2.27, 0.17, 0.15)$ & \cite{Linares2018} \\
GW 190917 sec. & $AN(2.1, 0.91, 0.3)$ & \cite{GWTC21} \\
GW 190425 prim. & $AN(2, 0.37, 0.18)$ & \cite{GWTC2} \\
\\
\enddata
\tablecomments{In each line we identify the name of the compact object itself, its companion or the event that generated it; as well as the individual mass distribution employed and its respective source. $N(\mu,\sigma)$ indicates a normal distribution with mean $\mu$ and standard deviation $\sigma$, while $AN(\mu, \sigma_1, \sigma_2)$ an asymmetrical normal distribution with peak at $\mu$, standard deviation $\sigma_1$ above $\mu$ and standard deviation $\sigma_2$ below. We include both the default weight (DW) and equal weight (EW) OGLE-2011-BLG-0463 mass estimates by \citet{Lam2022}, further discussed in the text.}
\end{deluxetable*}

The masses of objects J2215 \citep{Linares2018}, J0952 \citep{Romani2022} and of the companions to the giants J05215658 \citep{Thompson2019} and V723 Mon \citep{Jayasinghe2021} are obtained by means of light curve modeling as well as astrometric measurements; \citet{Romani2022} adjust both their full light curve data and their ``trimmed'' data with removed outliers, of which we adopt the more conservative trimmed fit. The mass of the secondary involved in the merger that gave rise to event GW190814 is determined from waveform models \citep{Abbott2020a}, while the mass of the remnant from GW170817 relies on waveform modeling as well as ejecta mass estimates from kilonova models \citep{Shibata2019}. 

The 'Unicorn' black hole, V723 Mon's companion, has its nature as a compact object challenged by \cite{2022MNRAS.512.5620E}, who conclude that it is actually a sub-giant. We perform our analysis both including and excluding this object, and verify whether our main conclusions change with its removal from the candidate pool.

OGLE-2011-BLG-0463, OB110462 for short, is the first definitive discovery of a compact object through astrometric microlensing, but shows considerable tension between the astrometric (favoring lower masses) and photometric measurements (favoring higher masses). \cite{Lam2022} consider two models: default weighting (DW), in which astrometric and photometric data are weighted according to the number of available observations, favoring photometric measurements; and equal weighting (EW), in which both are given the same weights. For the sake of definiteness, here we adopt DW as our fiducial model, but also discuss the EW case in Section \ref{sec:results}; both are shown in Table \ref{tab:newobjects}.
An independent analysis by \cite{Sahu_2022} yielded a mass of $7.1 \pm 1.3 M \odot$ that would put OB110462 outside the gap. Due to the discrepancy between \cite{Lam2022} and \cite{Sahu_2022}, \cite{https://doi.org/10.48550/arxiv.2207.10729} performed a third analysis that concluded that the discrepancy was due to systematic errors, settling the mass at $7.88 \pm 0.82 M \odot$. Since the results from \cite{Sahu_2022} and \cite{https://doi.org/10.48550/arxiv.2207.10729} result in OB110462 not being a gap candidate, we check for variation in our results when it is removed from the sample, as we did with V723 Mon.

Concerning uncertainties, we sought to model each mass distribution as either a Gaussian or two half-Gaussians with same $\mu$ but different $\sigma$ concatenated at the mean, which we call an asymmetrical Gaussian. For OB110462, J2215+5135 and GW190814, these distributions are shown to behave reasonably well as Gaussians in their respective sources (Table \ref{tab:newobjects}). For V723 Mon we were able to recover the distribution for $M_\mathrm{comp}$ and to determine that it is well-fitted by a Gaussian by applying Equation 4 to the \texttt{PHOEBE} results from Table 4 in that work. For the rest, distributions were assumed to be Gaussian. Uncertainties, whenever not given as $1\sigma$, were converted to and standardized as $1\sigma$.

\section{Analysis}
\label{sec:methods}

Given that the 12 objects described in Section \ref{sec:gap_objects} are mostly clustered near the lower edge of the gap, our analysis paid particular attention to this region, i.e. the $2-3\,\Msun$ interval, which the measurements can more significantly constrain, and the general features of the postulated mass gap. We describe in the following the statistical tests employed.

\subsection{Joint probability and Cumulative Distribution Function tests}
\label{subsec:met_freq}

The most direct way of considering the impact of these measurements from a frequentist analysis is to simply ask what is the probability that each individual object might in fact have a mass within a certain range. Taking the distributions $\mathcal{D}_i$ in Table \ref{tab:newobjects} as the mass PDFs for each candidate, their individual probabilities are computed by integrating them over the expect range.
        
A measure of the probability of the existence of a desert gap in itself then requires us to ask: given $N$ measured $\mathcal{D}_i$, plus the $111$ observations underlying $\mathcal{D}_\mathrm{desert}$, what is the probability that none of them corresponds to a gap object? A quantitative answer is given by

\begin{equation}
    \label{eq:gap_prob}
    P(\mathrm{desert\,gap})= \left[1-\int_\mathrm{gap}\frac{1}{111+N}\left(111 \mathcal{D}_\mathrm{desert}\,+ \sum_{i=1}^{N}\mathcal{D}_i\right)\,\mathrm{d}M\right]^{111+N}.
\end{equation}

Since the lower edge of the gap is not well constrained, it is interesting to allow it to vary from $2$ to $4\,\Msun$ while keeping the upper limit of the integration interval at $5\,\Msun$. By comparing the resulting probability with and without the gap candidates, we measure how certain is the existence of a desert gap that extends below $4\,\Msun$, for a given lower bound. 

In addition to these methods on mass distributions, we also compare the CDFs with and without the gap candidates in order to evaluate whether the addition of the new objects to the population makes it incompatible with the joint, desert gap, distribution. This \textit{CDF comparison test} is performed as a Kolmogorov-Smirnov (KS) test over synthetic mass samples drawn from the two CDFs, with the null hypothesis that both samples were drawn from the same distribution. In addition to the KS statistic, $D_\mathrm{KS}$, we compute also the $p$-value for obtaining that $D_\mathrm{KS}$ if the null hypothesis were true, for each sample pair. We reject the null hypothesis if $D_\mathrm{KS}$ is greater than the KS critical value for a level-of-significance of $\alpha=5\%$, with $p<5\%$.

\subsection{Likelihood ratio test}
\label{subsec:met_bay}

For a more refined evaluation of the depleted gap model, we compare the likelihood of the gap candidate sample, together with the underlying population from the desert gap model  ($\mathcal{D}_\mathrm{desert}$), having been drawn from $\mathcal{D}_\mathrm{desert}$ to that of being drawn from a depleted gap model ($\mathcal{D}_\mathrm{depleted}$). In order to keep our approach as simple as possible, we build our depleted model as the joint distribution from Section \ref{sec:mass_distr} with a \textit{plateau} between the NS and BH peaks (Figure \ref{fig:plateau}). We allow the height $h$ of this plateau to vary, and continuity determines its lower and upper bounds, $M^\mathrm{plat}_\mathrm{lower}$ and $M^\mathrm{plat}_\mathrm{upper}$, where it intersects the joint distribution; in this way the area below the plateau is also determined. The number $n$ of objects falling in the plateau is then

\begin{equation}
\label{eq:n_def}
    n = h\left(M^\mathrm{plat}_\mathrm{upper}-M^\mathrm{plat}_\mathrm{lower}\right)\left(111+N\right),
\end{equation}

\noindent where $N$ is again the number of gap candidates considered. We note that there is a maximum plateau height $h$ in this model set by the BH peak, so that the plateau cannot extend beyond it (see Figure \ref{fig:plateau}). For different values of $n$, we compute the likelihood ratio (LR) between $\mathcal{D}_\mathrm{desert}$ and $\mathcal{D}_\mathrm{depleted,n}$ as

\begin{equation}
\label{eq:bayes_factor}
    \mathrm{LR}(n) = \frac{\mathcal{L}\left(M|\mathcal{D}_\mathrm{depleted,n}\right)}{\mathcal{L}\left(M|\mathcal{D}_\mathrm{desert}\right)} = \frac{\prod^{111+N}_{i=1}P_\mathrm{depleted,n}(M_i)}{\prod^{111+N}_{i=1}P_\mathrm{desert}(M_i)},
\end{equation}

\noindent where $\mathcal{L}\left(M|\mathcal{D}\right)$ is the likelihood of model $\mathcal{D}$ describing the data points $M$, here the $111+N$ mass measurements $M_i$; and $P_\mathcal{D}(M_i)$ is the probability of $M_i$ being measured given the model $\mathcal{D}$. Note that $\mathcal{D}_\mathrm{depleted,n}$ reduces to $\mathcal{D}_\mathrm{desert}$ for $n=0$. 

Because we are working only with their fitted distributions, and not with the raw 16 BH and 95 NS data points, the 111 $M_i$ corresponding to these objects are drawn from their joint distribution. In order to take into account the variation in the mass measurements of our 12 gap candidates, their $M_i$ are also drawn, one from each distribution. To compensate for the small sample size, we draw multiple $M_i$ samples and compute $\mathrm{LR}(n)$ for each one of them. Because some variation in the $n$ for which $\mathrm{LR}$ peaks can occur between draws, we take the mean of this peak $n$ over draws as the preferred $n$.

\section{Results}
\label{sec:results}

\subsection{Joint probability and Cumulative Distribution Function results}
\label{subsec:res_freq}

\begin{figure}[h!]
\plotone{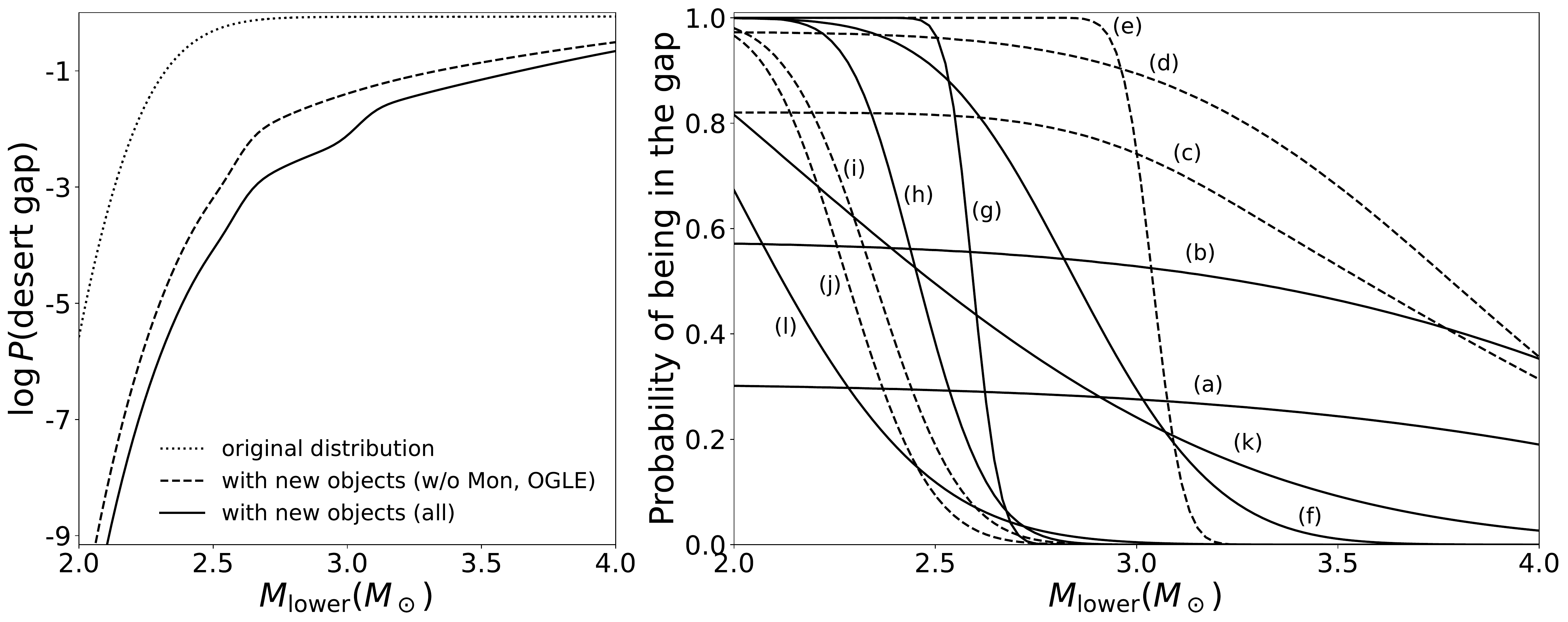}
\caption{Left: probability of existence of a desert gap with upper limit $5\,\mathrm{M}_\odot$ as a function of its lower limit, for the mixed distribution from Section \ref{sec:mass_distr} on its own (dotted line), including all gap candidates (solid line) and all gap candidates but V723 Mon and OB110462 (dashed line). Right: probability of each object falling in a gap with upper limit $5\,\mathrm{M}_\odot$ as a function of its lower limit; the objects are labeled as, from the most to the least massive: GW200115 (a), GW190924 (b), OB110462 (c), J02515658 (d), V723 Mon (e), GW200210 (f), GW190814 (g), GW170817 (h), J0952 (i), J2215 (j), GW190917 (k), GW190425 (l). Galactic sources are plotted with a dashed line, and extragalctic with a full line.}
\label{fig:gap_probabilities}
\end{figure}

In Figure \ref{fig:gap_probabilities} we show the results of the probability calculated with Equation \ref{eq:gap_prob}. On the left we consider three cases: the original distribution on its own, the original distribution plus the 12 new gap candidates and the original distribution plus all candidates but V723 Mon and OB110462. By keeping the gap upper limit at $5\,\Msun$ and varying the lower limit $M_\mathrm{lower}$ between $2$ and  $4\,\Msun$, we find that in every case the new candidates reduce the probability of a desert gap by at least $1$ order of magnitude below $3\Msun$. Comparing first with the full gap sample, starting from $M_\mathrm{lower}=2\,\Msun$, the probability of an absolute gap reaches $1\%$ at $2.21\,\Msun$ for the original distribution, but only at $2.85\,\Msun$ when considering the gap candidates. For the original model, the desert gap probability reaches $10\%$ for a $2.33-5\,\Msun$ gap, about $78\%$ between $3-5\,\Msun$ and a maximum of $86\%$ between $4-5\,\Msun$. With the gap candidates taken into account, the probability only reaches $\approx2\%$ for a $3-5\,\Msun$ gap, $10\%$ only for a $3.4-5\,\Msun$ gap and a maximum of $25\%$ for a $4-5\,\Msun$ gap.

Because they are some of the strongest candidates, the removal of V723 Mon and OB110462 cause a considerable difference in the probabilities, visible in the comparison between the solid and dashed lines in Figure \ref{fig:gap_probabilities}. V723 Mon, which provides a very narrow measurement well within the gap region, is responsible for the step-like feature visible in the solid line, which disappears in the dashed line; while OB110462, which has a broad distribution around the center of the gap, once removed causes an overall increase in the probabilities of there being a gap. These remain very low, however: considering the remaining 10 candidates, probability $1\%$ is reached for $M_\mathrm{lower}$ at $2.56\,\Msun$, close to the $M_\mathrm{max}^\mathrm{NS}=2.59\,\Msun$ found by \cite{Rocha2021}; $10\%$ at $2.99\,\Msun$; and a maximum of $36\%$ for a $4-5\,\Msun$ gap. Therefore, even after dismissing both candidates the evidence against a desert gap is still quite strong.

On the right side of Figure \ref{fig:gap_probabilities} we see the probabilities of each individual candidate falling within a gap between $M_\mathrm{lower}$ and $5\,\Msun$. With exception of GW200115, GW190924 and GW190425, these are greater than $66\%$ up to $M_\mathrm{lower}=2.2\,\Msun$; and for all objects they are greater than $10\%$ up to $M_\mathrm{lower}=2.5\,\Msun$. For the strongest candidates in our sample - OB110462 ($\approx3.79\,\Msun$), J05215658 ($\approx3.3\,\Msun$) and V723 Mon ($\approx3.04\,\Msun$) -, the probability of being gap objects never falls below $70\%$ while $M_\mathrm{lower}<3\,\Msun$, reaching at $3\,\Msun$ around $89\%$, $74\%$ and $75\%$, respectively. The next strongest candidate is an extragalactic source, the secondary from GW190924, with $52\%$ probability of being in the gap at $3\,\Msun$. Also shown object-wise in Table \ref{tab:biases} is the fractional decrease of the probability of there being a desert gap with $5\,\Msun$ as upper limit and four different values of $M_\mathrm{lower}$. With the exception of GW 200115, all objects cause a severe decrease in the probability of a desert gap that starts at or below $2.2\,\Msun$. A desert gap starting at or below $3\,\Msun$, on the other hand, is disfavored by those measurements that reach into the upper regions of the gap: GW190917, GW200210, V723 Mon, J05215658, OB110462, GW190924 and GW200115. The strongest evidence against the gap comes from V723 Mon and OB110463, if they are gap compact objects; and from J05215658 and GW190924 otherwise.

\begin{deluxetable*}{lcccc}[ht!]
\tabletypesize{\scriptsize}
\tablewidth{0pt} 
\tablecaption{Effect of Individual Gap Candidates on the Probability of a Gap for Varying Lower Limits \label{tab:biases}}
\tablehead{
\colhead{Name} & \multicolumn{4}{c}{$P\left(\mathrm{desert\, gap}|\mathrm{candidate}\right)/P\left(\mathrm{desert\, gap}\right)$} \\
{ } & \multicolumn{4}{c}{$M_\mathrm{lower}$} \\
{ } & $2.0\,\Msun$ & $2.2\,\Msun$ & $2.5\,\Msun$ & $3.0\,\Msun$
} 
\startdata 
GW 200115 prim. & 0.72 & 0.73 & 0.74 & 0.76 \\
GW 190924 sec. & 0.53 & 0.55 & 0.57 & 0.59 \\
OGLE-2011-BLG-0463 (DW) & 0.34 & 0.36 & 0.38 & 0.41 \\
OGLE-2011-BLG-0463 (EW) & 0.48 & 0.58 &  0.71 & 0.89\\
2MASS J05215658+4359220 comp. & 0.40 & 0.42 & 0.44 & 0.47 \\
V723 Mon comp. & 0.33 & 0.35 & 0.36 & 0.47 \\
GW 200210 sec. & 0.33 & 0.35 & 0.40 & 0.74 \\
GW 190814 sec. & 0.33 & 0.35 & 0.38 & 1.00 \\
GW 170817 rem. & 0.33 & 0.36 & 0.68 & 1.00 \\
PSR J0952-0607 & 0.33 & 0.43 & 0.83 & 1.00 \\
PSR J2215+5135 & 0.34 & 0.48 & 0.91 & 1.00 \\
GW 190917 sec. & 0.40 & 0.49 & 0.61 & 0.78 \\
GW 190425 prim. & 0.47 & 0.66 & 0.89 & 1.00 \\
\\
\enddata
\tablecomments{Given the joint distribution described in Section \ref{sec:mass_distr}, we compute the probability $P\left(\mathrm{desert\, gap}\right)$ of a desert gap between an upper limit $5\,\Msun$ and a varying lower limit taking on the values $2.0$, $2.2$, $2.5$ and $3.0\,\Msun$. In each line, we consider each gap candidate and again compute the probability of a desert gap in each interval, given both the joint distribution and the candidate, $P\left(\mathrm{desert\, gap}|\mathrm{candidate}\right)$. On the four right columns we show the ratio $P\left(\mathrm{desert\, gap}|\mathrm{candidate}\right)/P\left(\mathrm{desert\, gap}\right)$, that is, how much less probable a desert gap in that range becomes due to the individual object.}
\end{deluxetable*}

Given that the microlensing event OB110462 stands as the strongest gap candidate in the sample, we take into account the uncertainty in its mass estimate expressed by \citet{Lam2022} in considering also the EW model, in which the lens mass drops sharply, from $3.79^{+0.62}_{-0.57}\,\Msun$ in the DW model to $2.15^{+0.67}_{-0.54}\,\Msun$. By adopting this second estimate, the probability of the microlensing event being in the gap falls from $65\%$ to $11\%$ as $M_\mathrm{lower}$ goes from $2$ to $3\,\Msun$. However, even in this case, the total probability of there being a desert gap can still only reach a maximum of $\approx5\%$, surpassing $\approx10\%$ only when $M_\mathrm{lower}>3.1\,\Msun$. Thus even in this case a desert gap with lower bound below $3\,\Msun$ is strongly disfavored. 

\begin{figure}[ht!]
\plotone{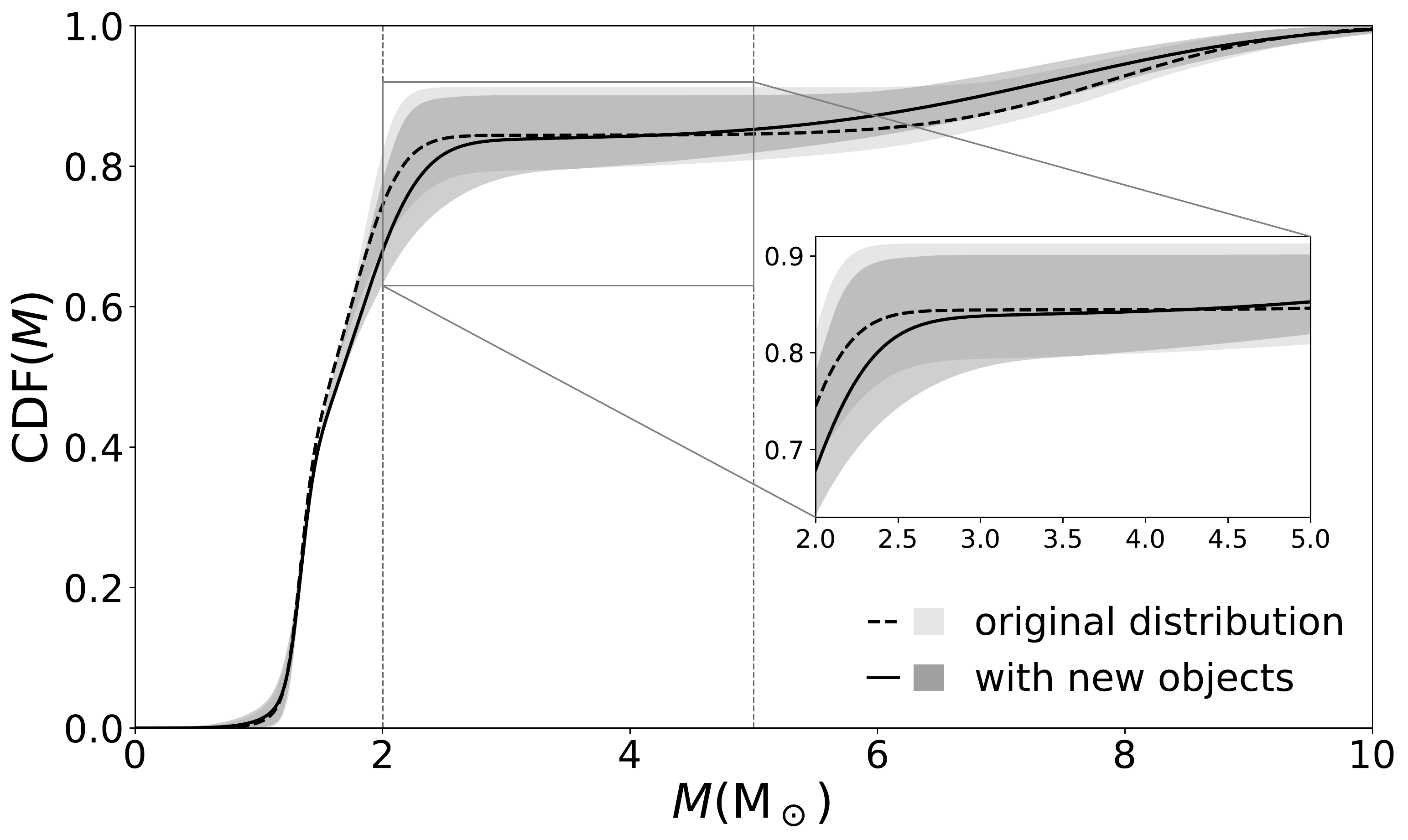}
\caption{Cumulative distribution functions (CDFs) for the original mixed mass distribution (dashed line) and the mixed distribution plus the new gap objects (solid line), with the gap range delimited by the two vertical dashed lines. The shaded regions represent the interval between the maximal and minimal draws for the original mixed distribution alone (lighter shade) and the one with the new objects (darker shade). While the variation of the CDF is minimal in the NS range of $\lesssim1.8\,\Msun$, it becomes significant in the gap range, specially in the $2-3\,\Msun$ interval, as can be more clearly seen in the inset. A pronounced increased in the CDF is also seen in the BH range beyond the gap, as a result of a greater proportion of objects laying within the gap region. A quantitative evaluation of these differences is discussed in the text.
\label{fig:cdfs}}
\end{figure}

In Figure \ref{fig:cdfs} we show the cumulative distribution functions (CDFs) for the original mixed mass distribution and the mixed distribution complemented by the 12 gap candidates. The variation in the CDF caused by the new objects is most visible around the lower edge of the gap region; it is seen in the inset that a suppression of the CDF extends up to $\sim3\,\Msun$, indicative of support for at most a depleted gap in the range, which agrees with previous discussion. In the inset of Figure \ref{fig:cdfs} it is seen that the presence of objects in the gap also causes a slight increase in the CDF around its upper edge, while the middle shows little to no variation. The further increase in the CDF above the gap is a consequence of the significant probability of the presence of objects somewhere within the gap.

A quantitative global measurement of the significance of these variations is obtained by performing the CDF comparison test as described in Section \ref{subsec:met_freq}. We draw $2000$ masses from each CDF and perform the test over the pair, then repeat the process for $10^6$ times. We find the mean values $D_\mathrm{KS}=0.08$ and $p=0.06\%$. For this sample size and $\alpha=5\%$ the critical KS value is $0.043$ and we can thus conclude that the full sample is incompatible with the desert gap model to a $95\%$ level of confidence. The same procedure, with V723 Mon and OB110462 removed results in $D_\mathrm{KS}=0.07$ and $p=0.2\%$, so that our result is robust even with the remove of the two candidates. A broader comparison of the desert and depleted gap models can be made with a likelihood ratio test, as shown in the next section.

\subsection{Likelihood ratio results}
\label{subsec:res_bay}

\begin{figure}[h!]
\plotone{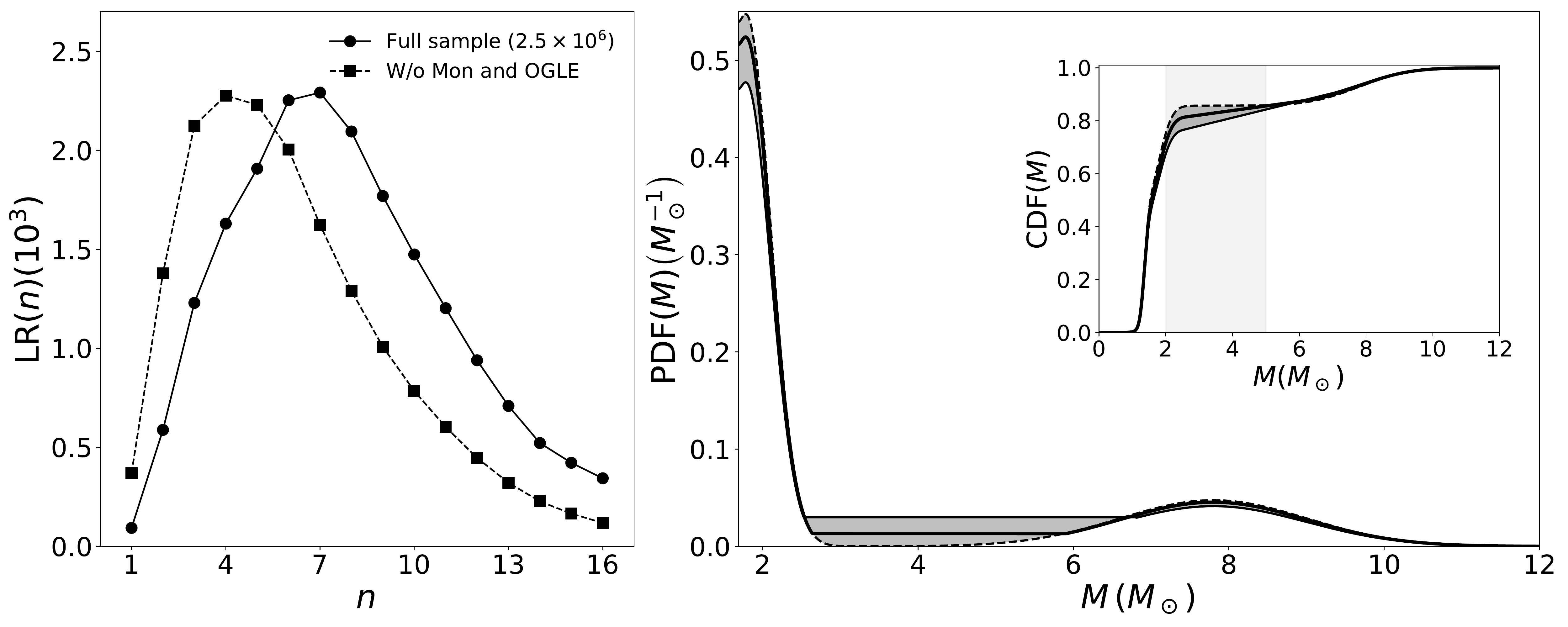}
\caption{Left: resulting likelihood ratio from Equation \ref{eq:bayes_factor} for varying values of $n$, for the full gap candidate sample (solid line) and with V723 Mon and OB110463 removed (dashed line). Both curves show a typical draw that results in a peak closest to the mean preferred $n$ over random draws. The likelihood ratio peaks at $n=7$ for the full sample and at $n=4$ with those two candidates removed. Note that the solid line is divided by a factor of $2.5\times10^6$ to fit the scale with the dashed line, which is nonetheless still significant. Right: the PDF and the CDF for models varying between $n=0$ (dashed line, desert gap) and $n=16$ (thin solid line). The central thick solid line shows the preferred distribution of a depleted gap with $n=7$. Notice that there is a maximum allowed height for the plateau so that it does not extend beyond the BH peak.
\label{fig:plateau}}
\end{figure}

We computed the likelihood ratio $\mathrm{LR}(n)$ between the $\mathcal{D}_\mathrm{desert}$ and $\mathcal{D}_\mathrm{depleted,n}$ models for $n$ varying from $1$ to $16$, in relation to the full sample of $N=123$ objects. We show the results for two cases on the left side of Figure \ref{fig:plateau}, with the full sample as the solid line; the solid line is shown divided by a factor of $2.5\times10^6$ to fit the scale of the figure for comparison with the reduced sample (dashed line, described below). For all $n$ tested, the depleted gap model was strongly preferred over the desert gap model. Although we do not claim the plateau to describe the actual shape of the physical compact object mass distribution, this strong preference for even the simplest non-desert gap model indicates that there cannot be a desert gap to a high level of confidence.

In terms of the depth of this gap, and whether there could still be a depletion zone within $2-5\,\Msun$, we do find a clear preference for a model with $n=7$, with $M^\mathrm{plat}_\mathrm{lower}=2.61\,\Msun$ and $M^\mathrm{plat}_\mathrm{upper}=6.12\,\Msun$. The existence of a peak on the left side of Figure \ref{fig:plateau} is expected when testing the models against the full sample, because this procedure creates a preference for a model that reflects the proportion between objects in and out of the gap in the sample, instead of allowing $\mathrm{LR}(n)$ to grow indefinitely with $n$. Although it might be expected that $n=12$ would be preferred, even this model would place the lower limit as far up as $M^\mathrm{plat}_\mathrm{lower}=2.56\,\Msun$, while most candidates from Table \ref{tab:newobjects} have nominal masses below that value; on the other end, our most massive candidates reach into the BH peak, such that the plateau does not have to fully account for them. Thus it is not surprising that the preferred model has $n<12$.

On the right side of Figure \ref{fig:plateau} we show the variation of the distribution from model $\mathcal{D}_\mathrm{depleted,n}$ from $n=0$ ($\mathcal{D}_\mathrm{desert}$) to $16$ in the mass PDF and CDF, with the preferred $n=7$ model as the central distribution. As this indicates, even with the addition of the recent gap candidates, a depletion in the gap range is still expected over the complete absence of a gap, which for our model would result in very high values of $n$ being preferred. As with the tests of Section \ref{subsec:res_freq}, a depleted gap is again preferred over a desert gap.

We performed the same procedure after removing Mon V723 and OB110462 from the sample, resulting in the dashed line on the left side of Figure \ref{fig:plateau}. We found that the depleted gap is still preferred over a desert overall, but with a LR peak at $n=4$, yielding a plateau between $M_\mathrm{lower}=2.66\,\Msun$ and $M^\mathrm{plat}_\mathrm{upper}=5.77\,\Msun$. As in Section \ref{subsec:res_freq}, removal of these two candidates is significant, but our results remain robust.

\section{Conclusions}
\label{sec:conclusion}

In this work we tested the existence of a desert mass gap between the NS and BH mass distributions against recently observed compact objects with well-measured masses to varying degrees. Such a gap has long been proposed to exist between $2$ and $5\,\Msun$ \citep{Bailyn1998, Ozel2010, Farr2011}, starting just above the maximum NS mass. Although predictions for this maximum mass have reached higher values, up to $2.6\,\Msun$ \citep{Margalit_2017,Ruiz2018,Alsing2018,Rezzolla2018,Shibata2019,Ai_2020,Shao2020,Rocha2021}, and no conclusive theoretical reason other than the formation process has been established to support such a break in the compact object mass distribution \citep{Fryer_2012, Belczynski_2012, Zevin2020, Liu2021, Fryer2022, Olejak2022}, the idea that such a feature might exist has persisted, in spite of the result that systematic overestimations might be present in BH mass measurements \citep{Kreidberg2018}. With an ever-growing catalog of compact object mass measurements, however, the $2-5\,\Msun$ range has started to be populated, and the desert gap model can now be directly explored. 

As shown in Section \ref{subsec:res_freq}, the recently observed candidate gap objects strongly disfavor a desert gap starting below $3\,\Msun$. For our default sample of candidates, the probability of a desert gap reaching or extending below $3\,\Msun$ is shown to be at most $\approx2\%$. We also consider the possibility that V723 Mon and OB110462 might not at all be gap compact objects, but even in this case the probability of a desert $3-5\,\Msun$ gap increases only to $\approx10\%$, and remains smaller than $1\%$ for a gap starting at $2.56\,\Msun$ or below. When considering the lower, equal-weight mass estimate for the microlensing event OB110462, the $3-5\,\Msun$ probability only reaches up to $\approx5\%$. By looking at the individual measurements, V723 Mon, J05215658 and GW190924 prove to be strong indicators that there cannot be a desert gap, with respective probabilities of at least $75\%$, $74\%$ and $52\%$ of being located within a gap that starts anywhere between $2-3\,\Msun$ and goes up to $5\,\Msun$. On the other hand, OB110462 is, in the default weight model, the strongest candidate with a probability of at least $89\%$ of being inside such a gap for any value of its lower limit. By performing a CDF comparison test between the Gaussian mixture of the NS and BH mass distributions by \cite{Rocha2021} and \cite{Ozel2010}, respectively, and the Gaussian mixture of those distributions and those of our candidates, we have found that there is only a $p=0.06\%$ probability of the two population ensembles being drawn from the same underlying distribution. Discarding V723 Mon and OB110462 only increases this value to $p=0.2\%$, indicating that the recent measurements are incompatible with a desert gap to a high degree of confidence.

We also performed a likelihood ratio test to compare the desert and depleted gap models by treating the depletion zone as a plateau with a varying height determined by the number $n$ of objects in that region, such that $n=0$ corresponds to the desert gap model. By computing the likelihood ratio, LR, between the desert and depleted models for $n$ up to $16$, we found that a depleted model is always strongly preferred over the desert model, with the overall preferred model being that for $n=7$ objects in a gap between $2.61$ and $6.12\,\Msun$. While we stress that we do not suggest that a plateau is an accurate representation of the mass distribution in the gap, the strong odds obtained for this simple depleted gap model do indicate that there is no desert gap. The fact that LR clearly falls for $n>7$ also points towards there still being a gap, albeit not an empty (desert) one. While LR drops a few orders of magnitude when Mon V723 and OB110462 are discarded, the same behavior is observed, with the preferred plateau being that with $n=4$.

It is also worth pointing out that the 12 candidate objects included here do {\it not} exhaust the pool of observed candidate gap objects. While our sample includes only those measurements deemed as stronger or more reliable candidates, other candidates include 1E 1740.7-2942, a hard X-ray emitter suggested to have a mass between $4-5\,\Msun$ by \citet{Stecchini2020}; the remnant of the binary NS merger GW190425, with a total mass of only $3.04^{+0.3}_{-0.1}\,\Msun$, but no observed electromagnetic counterpart \citep{Abbott2021a}; PSR J1311-3430, with mass in the $1.8-2.7\,\Msun$ range \citep{Romani2015}; and PSR J1748-2021B, with mass in the $2.1-2.9\,\Msun$ range \citep{Freire2008, OzelFreire2016}, currently scheduled for an attempt at observing its binary companion with the James Webb Space Telescope, which, if detected, should allow stricter constraints to be placed on the mass of the pulsar \citep{Freire2021JWST}. As these observations accumulate, we can expect even stronger constraints to be placed on the existence of the gap, its character as a depletion zone, and on its range and depth, to be understood together with the events that contribute to it. Although we have taken into account here recent BH mass measurements that are in or skirt the gap range, a full appreciation of the extent to which the gap has been filled will require an updated BH mass distribution (Bernardo et al., in preparation). We list the mass measurements of 8 additional, weaker or less precise, gap candidates in Appendix \ref{app:additional_gap}.

While we have sought to show that current observational evidence is enough to favor the existence of a depleted, but not desert, gap, we do not attempt to define the particular shape that the distribution should take in order to accommodate this depletion gap zone. This question has been recently addressed by \citet{Farah2022}, who fitted a broken power-law model with a dip representing the gap to mass measurements from gravitational wave data. They found a gap between $2.2^{+0.7}_{-0.5}\,\Msun$ and $6.0^{+2.45}_{-1.4}\,\Msun$, which encompasses both the $n=7$ and $n=4$ results we find for a plateau model. For their ``depth parameter'' $A$, $A = 0$ indicating the absence of a gap and $A = 1$ a maximal gap, they found a peak value of $A=0.9$ with a $90\%$ confidence interval between $0.55$ and $1$, favoring the existence of a gap which might be only partial. As the authors suggest, it is enough for a few objects to have considerable support within the gap range for its inferred depth to be considerably decreased; in their case these were the secondaries of GW190814 ($P\left(\mathrm{in\;gap}\right)=74\%$) and GW190924\_021846 ($P\left(\mathrm{in\;gap}\right)=49\%$), and the primary of GW200115 ($P\left(\mathrm{in\;gap}\right)=44\%$), which we also found to be significant gap candidates, with GW190924 providing the strongest constraint on the gap from an extragalactic source.

With regard to the simple question of whether there can be a desert gap between the NS and BH mass distribution peaks, independently of the specific shape of the full compact object mass distribution and with minimal assumptions, we show that current observations strongly support that this gap is at least partially filled. It can thus be safely concluded that there is at most a depleted, but not a desert, mass gap in the full compact object distribution.

In terms of how this gap is partially filled, binary population synthesis may help to examine the role different compact object formation channels play. While a inefficiency or inability of core-collapse supernovae to produce gap objects has been linked to convection time scales and fallback amounts \citep{Zevin2020,Liu2021,Fryer2022,Olejak2022}, other process might still be able to play this ``gap filling'' role, such as the accretion induced collapse of NSs into BHs in redback/black widow binaries \citep{Horvath2020}; NS-NS mergers \citep{Gupta2020}, of which the Galaxy should harbor $\sim 10^{4}$ remnants using the simplest merger rate estimate; or the double-degenerate accretion induced collapse of white dwarfs into massive NSs \citep{Wang2020}. The relative contribution of each one remains to be investigated in full.

\begin{acknowledgments}
ACKNOWLEDGMENTS

We would like to acknowledge an anonymous referee and the editorial staff for a careful appreciation of the manuscript and several suggestions that helped improve the final version. We also thank Dorota Rosinska for scientific advice. The authors wish to acknowledge the financial support of the Fapesp Agency (S\~ao Paulo) through the grant 13/26258-4 and 2020/08518-2 and the CNPq (Federal Government) for the award of a Research Fellowship to JEH. LMS acknowledges CNPq for financial support. The CAPES Agency (Federal Government) is acknowledged for financial support in the form of Scholarships.

\textit{Software:}  \texttt{PyMC3} \citep{PyMC3}, \texttt{NumPy} \citep{Numpy}, \texttt{SciPy} \citep{Virtanen_2020}, \texttt{Matplotlib} \citep{matplotlib},  \textit{Mathematica} \citep{Mathematica}.

\end{acknowledgments}

\clearpage

\appendix

\section{Additional gap candidates}
\label{app:additional_gap}

In addition to the 12 gap candidates studied in this work, there exists a broader collection of potential gap objects that are either weaker candidates or have thus far not received accurate enough mass measurements for the methods we have employed. We show here the full list of 20 objects we took into account, of which 8 were not included in our analysis. These 8 include the secondaries of events GW190930, GW200316, GW190725 and GW191113, BHs with mass distributions that do not reach the $2-5\,\Msun$ range within $1\sigma$, but do within their $90\%$ confidence intervals; event GW190425, with a total mass in the gap but no remnant mass estimate due to the lack of an electromagnetic counterpart; the pulsars J1311-3430 and J1748-20212B, which so far have received only upper and lower mass bounds; and the hard X-ray emitter 1E 1740.7-2942, which has also only received a mass range.

\begin{figure}[ht!]
\plotone{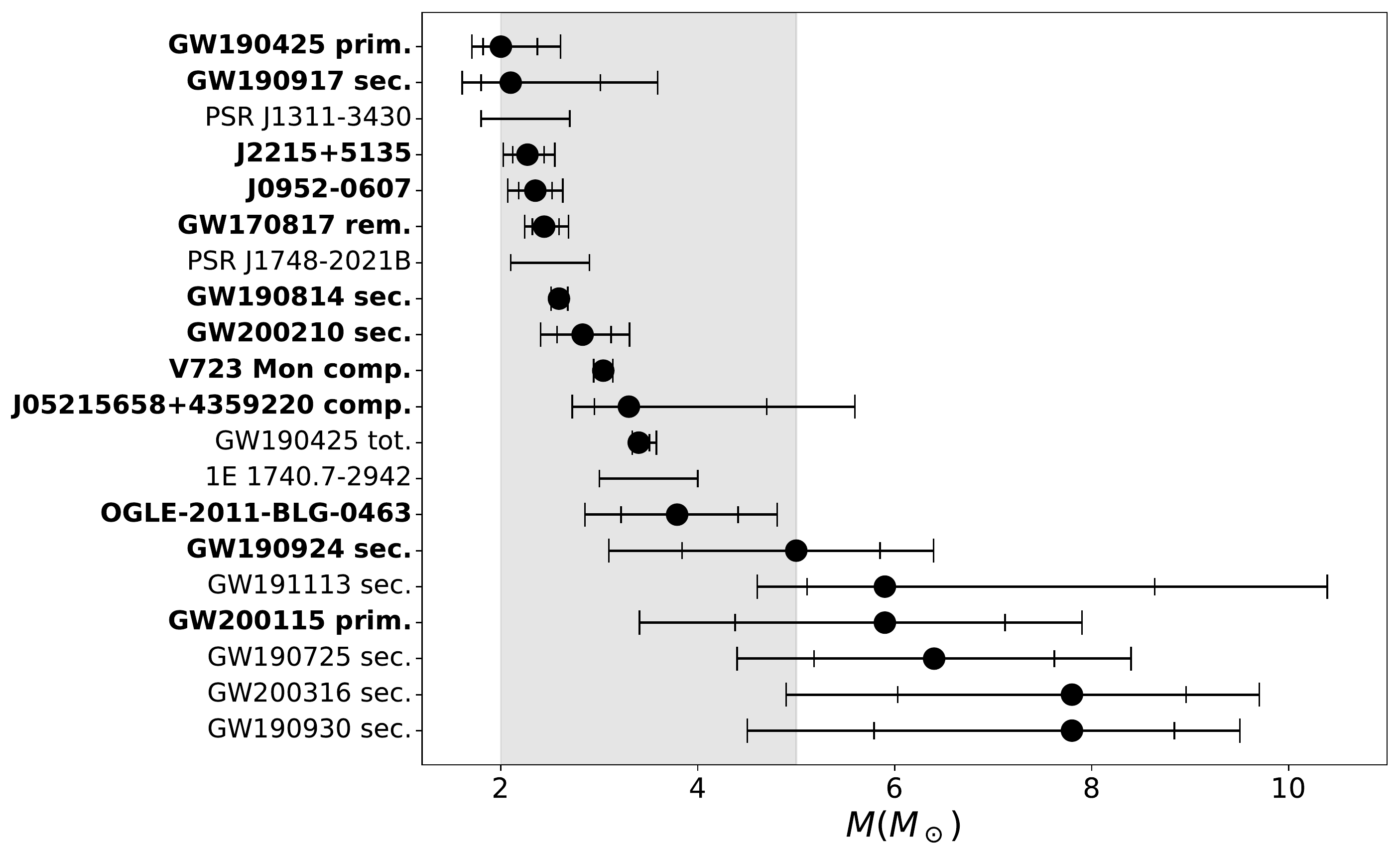}
\caption{Mass distributions for the 12 gap candidates considered in the preceding analysis (names in bold) and for 8 additional candidates. For PSR J1311-3430, PSR J1748-2021B and 1E 1740.7-2942 only the available mass ranges are shown, while for the other objects we show both nominal values, $1\sigma$ (inner bars) and $90\%$ (outer bars) confidence intervals. We highlight the gap range in gray. \label{fig:gapdotplot}}
\end{figure}

We display in Table \ref{tab:gaptab} the references and mass distributions or ranges, and in Figure \ref{fig:gapdotplot} the mass ranges, of the 12 candidates we studied, highlighted in bold, and of the 8 additional candidates. 

\begin{deluxetable*}{lcl}[h]
\tabletypesize{\scriptsize}
\tablewidth{0pt} 
\tablecaption{Mass Distributions for 20 Gap Candidates\label{tab:gaptab}}
\tablehead{
\colhead{Name} & \colhead{$M$}& \colhead{Reference}
\\
{} & $\left(\Msun\right)$ & {}} 
\startdata 
GW 190930 sec. & $AN(7.8, 1.7, 3.3)$ & \cite{GWTC2} \\
GW 200316 sec. & $AN(7.8, 1.9, 2.9)$ & \cite{GWTC3} \\
GW 190725 sec. & $N(6.4,2)$ & \cite{GWTC21} \\
GW 191113 sec. & $AN(5.9, 4.5, 1.3)$ & \cite{GWTC3} \\
\textbf{GW 200115 prim.} & $AN(5.9, 2, 2.5)$ & \cite{GWTC3} \\
\textbf{GW 190924 sec.} & $AN(5, 1.4, 1.9)$ & \cite{GWTC2} \\
1E 1740.7-2942 & $4-5\,\Msun$ & \cite{Stecchini2020} \\
\textbf{OGLE-2011-BLG-0463 (DW)} & $AN(3.79, 0.62, 0.57)$ & \cite{Lam2022} \\
\textbf{OGLE-2011-BLG-0463 (EW)} & $AN(2.15, 0.67, 0.54)$ & \cite{Lam2022} \\
GW 190425 rem. & $AN(3.4, 0.18, 0.06)$ & \cite{GWTC2} \\
\textbf{2MASS J05215658+4359220 comp.} & $AN(3.3, 1.4, 0.35)$ & \cite{Thompson2019} \\
\textbf{V723 Mon comp.} & $N(3.04, 0.06)$ & \cite{Jayasinghe2021} \\
PSR J1748-2021B & $2.1-2.9\,\Msun$ & \cite{Freire2008, OzelFreire2016} \\
\textbf{GW 200210 sec.} & $AN(2.83, 0.48, 0.43)$ & \cite{GWTC3} \\
PSR J1311-3430 & $1.8-2.7\,\Msun$ & \cite{Romani2015} \\
\textbf{GW 190814 sec.} & $N(2.59, 0.05)$ & \cite{Abbott2020a} \\
\textbf{GW 170817 rem.} & $AN(2.44, 0.15, 0.12)$ & \cite{Shibata2019}\\
\textbf{PSR J0952-0607} & $N(2.35, 0.17)$ & \cite{Romani2022} \\
\textbf{PSR J2215+5135} & $AN(2.27, 0.17, 0.15)$ & \cite{Linares2018} \\
\textbf{GW 190917 sec.} & $AN(2.1, 1.5, 0.5)$ & \cite{GWTC21} \\
\textbf{GW 190425 prim.} & $AN(2, 0.6, 0.3)$ & \cite{GWTC2} \\
\\
\enddata
\tablecomments{In each line we identify the name of the compact object itself, its companion or the event that generated it; as well as the individual mass distribution employed and its respective source. $N(\mu,\sigma)$ indicates a normal distribution with mean $\mu$ and standard deviation $\sigma$, while $AN(\mu, \sigma_1, \sigma_2)$ an asymmetrical normal distribution with peak at $\mu$, standard deviation $\sigma_1$ above $\mu$ and standard deviation $\sigma_2$ below. We include both the default weight (DW) and equal weight (EW) OGLE-2011-BLG-0463 mass estimates by \citet{Lam2022}, as discussed in Section \ref{sec:gap_objects}. The 12 candidates employed in our full analysis are highlighted in bold.}
\end{deluxetable*}

\section{Neutron star and black hole distributions}
\label{app:bhnsdistr}

The neutron star mass distribution contains more features than a simple ``canonical'' $1.4 M_{\odot}$, widely accepted in the past \citep[i.e., ][]{clark2002}. At least two maxima are present, as found in \cite{valentim2011, kiziltan2013neutron, antoniadis2016millisecond}. Although the precise numbers differ according to the sample and other assumptions, the presence of a bimodal distribution is quite robust and always preferred to any form. We employed here the distribution obtained by means of a Bayesian analysis presented in \cite{Rocha2021}, consisting in two Gaussian peaks located at $1.365\,\Msun$ and $1.787\,\Msun$ with respective $\sigma$ $0.109\,\Msun$ and $0.314\,\Msun$, and weight $0.498$ for the first peak. The same analysis rendered a high maximum mass $M^\mathrm{NS}_\mathrm{max}=2.59\,\Msun$, which complies with the highest reported observations \citep{Romani2022}, implying that the gap, if real, starts at a value larger than the original $2 M_{\odot}$.

\begin{figure}[ht!]
\plotone{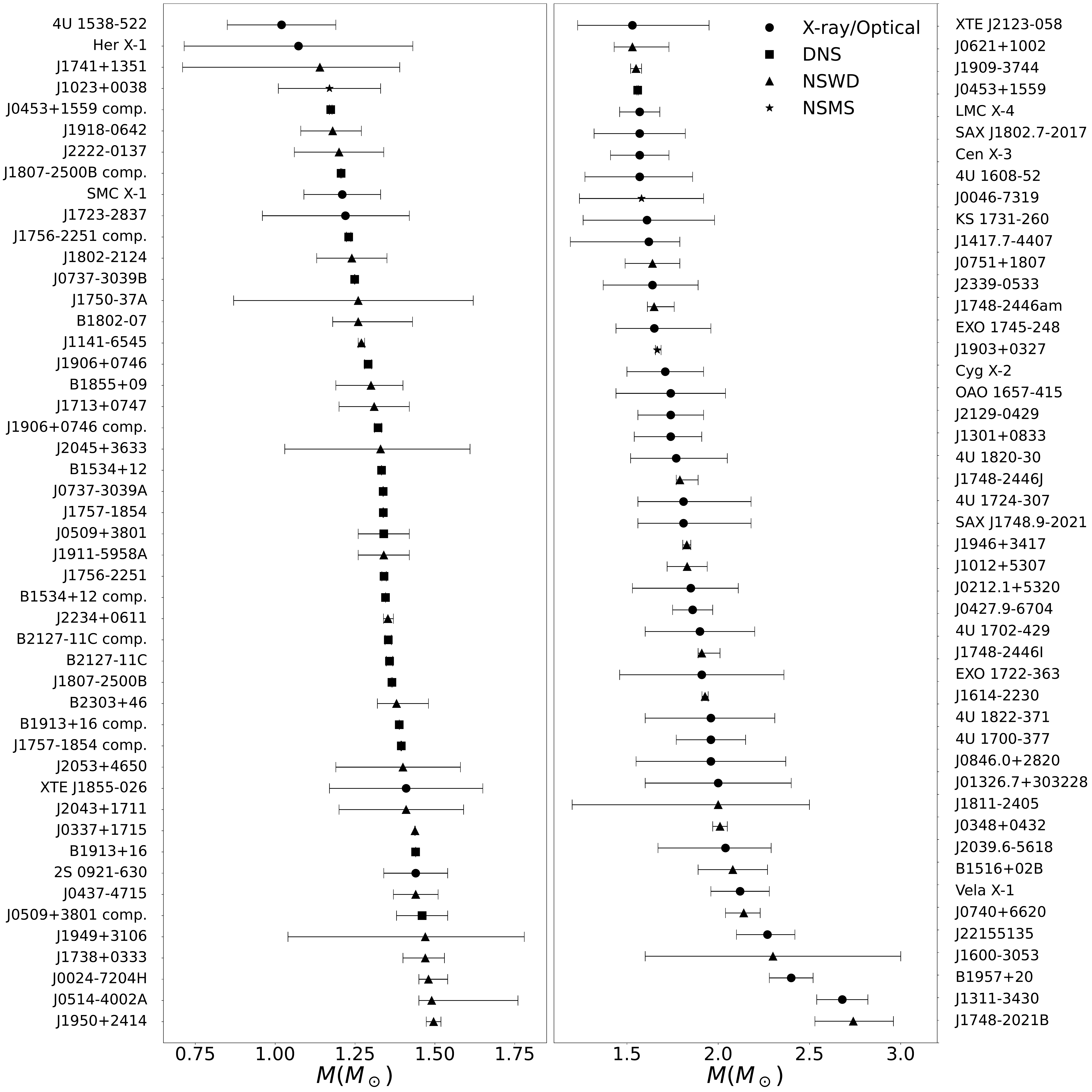}
\caption{Nominal values and $1\sigma$ credibility ranges for the 95 NS mass measurements from \cite{Rocha2021}, distinguished by system type. \label{fig:nsdotplot}}
\end{figure}

We show in Table \ref{tab:nstab} the mass measurements considered by \cite{Rocha2021} in adjusting their distribution. At the time of submission of this paper, \cite{Rocha2021} relate 96 masses with PSR B1855+09 being included twice. Removal of the duplicate affects minimally the distribution and does not change our results. These same measurements are also displayed in Figure \ref{fig:nsdotplot}.

\startlongtable
\begin{deluxetable*}{lccl}
\tabletypesize{\scriptsize}
\tablewidth{0pt} 
\tablecaption{Mass Measurements for 95 Neutron Stars from \cite{Rocha2021}\label{tab:nstab}}
\tablehead{
\colhead{Name} & \colhead{Type} & \colhead{$M$}& \colhead{Reference} \\
{} & {} & $\left(\Msun\right)$  & {}
}  
\startdata 
4U 1700-377 & X-ray/Optical & $1.96^{+0.19}_{-0.19}$ & \cite{falanga2015ephemeris} \\
Cyg X-2 & X-ray/Optical & $1.71^{+0.21}_{-0.21}$ & \cite{casares2010mass} \\
SMC X-1 & X-ray/Optical & $1.21^{+0.12}_{-0.12}$ & \cite{falanga2015ephemeris} \\
Cen X-3 & X-ray/Optical & $1.57^{+0.16}_{-0.16}$ & \cite{falanga2015ephemeris} \\
XTE J2123-058 & X-ray/Optical & $1.53^{+0.30}_{-0.42}$ & \cite{2002AAS...201.5405G}\\
4U 1822-371 & X-ray/Optical & $1.96^{+0.36}_{-0.35}$ & \cite{munoz2005k}\\
OAO 1657-415 & X-ray/Optical & $1.74^{+0.30}_{-0.30}$ & \cite{falanga2015ephemeris}\\
J01326.7+303228 & X-ray/Optical & $2.0^{+0.40}_{-0.40}$ & \cite{bhalerao2012neutron}\\
Vela X-1 & X-ray/Optical & $2.12^{+0.16}_{-0.16}$ & \cite{falanga2015ephemeris}\\
4U 1538-522 & X-ray/Optical & $1.02^{+0.17}_{-0.17}$ & \cite{falanga2015ephemeris}\\
LMC X-4 & X-ray/Optical & $1.57^{+0.11}_{-0.11}$ & \cite{falanga2015ephemeris}\\
Her X-1 & X-ray/Optical & $1.073^{+0.358}_{-0.358}$ & \cite{rawls2011refined}\\
2S 0921-630 & X-ray/Optical & $1.44^{+0.10}_{-0.10}$ & \cite{steeghs2007mass}\\
EXO 1722-363 & X-ray/Optical & $1.91^{+0.45}_{-0.45}$ & \cite{falanga2015ephemeris}\\
SAX J1802.7-2017 & X-ray/Optical & $1.57^{+0.25}_{-0.25}$ & \cite{falanga2015ephemeris}\\
XTE J1855-026 & X-ray/Optical & $1.41^{+0.24}_{-0.24}$ & \cite{falanga2015ephemeris}\\
B1957+20 & X-ray/Optical & $2.40^{+0.12}_{-0.12}$ & \cite{van2011evidence}\\
J1311-3430 & X-ray/Optical & $2.68^{+0.14}_{-0.14}$ & \cite{romani2012psr}\\
4U 1608-52 & X-ray/Optical & $1.57^{+0.30}_{-0.29}$ &  \cite{ozel2016dense}\\
4U 1724-307 & X-ray/Optical & $1.81^{+0.25}_{-0.37}$ & \cite{ozel2016dense}\\
EXO 1745-248 & X-ray/Optical & $1.65^{+0.21}_{-0.31}$ & \cite{ozel2016dense}\\
KS 1731-260 & X-ray/Optical & $1.61^{+0.35}_{-0.37}$ & \cite{ozel2016dense}\\
SAX J1748.9-2021  & X-ray/Optical & $1.81^{+0.25}_{-0.37}$ & \cite{ozel2016dense}\\
4U 1820-30 & X-ray/Optical & $1.77^{+0.25}_{-0.28}$ & \cite{ozel2016dense}\\
4U 1702-429 & X-ray/Optical & $1.90^{+0.30}_{-0.30}$ & \cite{Nattila}\\
J22155135 & X-ray/Optical & $2.27^{+0.17}_{-0.15}$ & \cite{Linares2018}\\
J1301+0833 & X-ray/Optical & $1.74^{+0.20}_{-0.17}$ & \cite{romani2016}\\
J0212.1+5320 & X-ray/Optical & $1.85^{+0.32}_{-0.26}$ & \cite{shabaz2017}\\
J1723-2837 & X-ray/Optical & $1.22^{+0.26}_{-0.20}$ & \cite{strader2019}\\
J1417.7-4407 & X-ray/Optical & $1.62^{+0.43}_{-0.17}$ & \cite{strader2019}\\
J2339-0533 & X-ray/Optical & $1.64^{+0.27}_{-0.25}$ & \cite{strader2019}\\
J2129-0429 & X-ray/Optical & $1.74^{+0.18}_{-0.18}$ & \cite{strader2019}\\
J0427.9-6704 & X-ray/Optical & $1.86^{+0.11}_{-0.11}$ & \cite{strader2019}\\
J0846.0+2820 & X-ray/Optical & $1.96^{+0.41}_{-0.41}$ & \cite{strader2019}\\
J2039.6-5618 & X-ray/Optical & $2.04^{+0.37}_{-0.25}$ & \cite{strader2019}\\
J0453+1559 & DNS & $1.559^{+0.004}_{-0.004}$ & \cite{martinez2015pulsar}\\
J0453+1559Cp & DNS & $1.174^{+0.004}_{-0.004}$ & \cite{martinez2015pulsar}\\
J1906+0746 & DNS & $1.291^{+0.011}_{-0.011}$ & \cite{van2015binary}\\
J1906+0746Cp & DNS & $1.322^{+0.011}_{-0.011}$ & \cite{van2015binary}\\
B1534+12 &  DNS & $1.3330^{+0.0002}_{-0.0002}$ & \cite{fonseca2014comprehensive}\\
B1534+12Cp & DNS &  $1.3455^{+0.0002}_{-0.0002}$ & \cite{fonseca2014comprehensive}\\
B1913+16 & DNS & $1.4398^{+0.0002}_{-0.0002}$ & \cite{weisberg2010timing}\\
B1913+16Cp & DNS & $1.3886^{+0.0002}_{-0.0002}$ & \cite{weisberg2010timing}\\
B2127-11C & DNS & $1.358^{+0.010}_{-0.010}$ & \cite{jacoby2006measurement}\\
B2127-11CCp & DNS & $1.354^{+0.010}_{-0.010}$ & \cite{jacoby2006measurement}\\
J0737-3039A & DNS & $1.3381^{+0.0007}_{-0.0007}$ &  \cite{kramer2006tests}\\
J0737-3039B & DNS & $1.2489^{+0.0007}_{-0.0007}$ &  \cite{kramer2006tests}\\
J1756-2251 & DNS & $1.341^{+0.007}_{-0.007}$ & \cite{2014MNRAS.443.2183F}\\
J1756-2251Cp & DNS &  $1.230^{+0.007}_{-0.007}$ & \cite{2014MNRAS.443.2183F}\\
J1807-2500B & DNS & $1.3655^{+0.0021}_{-0.0021}$ & \cite{lynch2012timing}\\
J1807-2500BCp & DNS &  $1.2064^{+0.0020}_{-0.0020}$ & \cite{lynch2012timing}\\
J0509+3801 & DNS & $1.34^{+0.08}_{-0.08}$ & \cite{lynch2012timing}\\
J0509+3801Cp & DNS & $1.46^{+0.08}_{-0.08}$ & \cite{lynch2012timing}\\
J1757-1854 & DNS & $1.3384^{+0.0009}_{-0.0009}$ & \cite{cameron2018}\\
J1757-1854Cp & DNS & $1.3946^{+0.0009}_{-0.0009}$ & \cite{cameron2018}\\
J2045+3633 & NS-WD & $1.33^{+0.30}_{-0.28}$ & \cite{berezina2017discovery}\\
J2053+4650 & NS-WD & $1.40^{+0.21}_{-0.18}$ & \cite{berezina2017discovery}\\
B1855+09 & NS-WD & $1.30^{+0.11}_{-0.10}$ & \cite{fonseca2016nanograv}\\
J1713+0747 & NS-WD & $1.31^{+0.11}_{-0.11}$ & \cite{zhu2015testing}\\
J0751+1807 & NS-WD & $1.64^{+0.15}_{-0.15}$ & \cite{desvignes2016high}\\
J1141-6545 & NS-WD & $1.27^{+0.01}_{-0.01}$ & \cite{bhat2008gravitational}\\
J1738+0333 & NS-WD & $1.47^{+0.07}_{-0.06}$ & \cite{antoniadis2012relativistic}\\
J1614-2230 & NS-WD & $1.928^{+0.017}_{-0.017}$ & \cite{fonseca2016nanograv}\\
J0348+0432 & NS-WD & $2.01^{+0.04}_{-0.04}$ & \cite{antoniadis2016millisecond}\\
J2222-0137 & NS-WD & $1.20^{+0.14}_{-0.14}$ & \cite{kaplan20141}\\
J2234+0611 & NS-WD & $1.353^{+0.014}_{-0.017}$ & \cite{stovall2019psr}\\
J1949+3106 & NS-WD & $1.47^{+0.43}_{-0.31}$ & \cite{deneva2012two}\\
J1012+5307 & NS-WD & $1.83^{+0.11}_{-0.11}$ & \cite{antoniadis2016millisecond}\\
J0437-4715 & NS-WD & $1.44^{+0.07}_{-0.07}$ & \cite{reardon2016}\\
J1909-3744 & NS-WD & $1.55^{+0.03}_{-0.03}$ & \cite{fonseca2016nanograv}\\
J1802-2124 & NS-WD & $1.24^{+0.11}_{-0.11}$ & \cite{ferdman2010precise}\\
J1911-5958A & NS-WD & $1.34^{+0.08}_{-0.08}$ & \cite{bassa2006}\\
J2043+1711 & NS-WD & $1.41^{+0.21}_{-0.18}$ & \cite{fonseca2016nanograv}\\
J0337+1715 & NS-WD & $1.4378^{+0.0013}_{-0.0013}$ & \cite{ransom2014millisecond}\\
J1946+3417 & NS-WD & $1.828^{+0.022}_{-0.022}$ & \cite{barr2017massive}\\
J1918-0642 & NS-WD & $1.18^{+0.10}_{-0.09}$ & \cite{fonseca2016nanograv}\\
J1600-3053 & NS-WD & $2.3^{+0.7}_{-0.7}$ & \cite{arzoumanian2018nanograv}\\
J0024-7204H & NS-WD & $1.48^{+0.03}_{-0.06}$ & \cite{kiziltan2013neutron}\\
J0514-4002A & NS-WD & $1.49^{+0.04}_{-0.27}$ & \cite{kiziltan2013neutron}\\
J0621+1002 & NS-WD & $1.53^{+0.10}_{-0.20}$ & \cite{kasian2012radio}\\
B1516+02B & NS-WD & $2.08^{+0.19}_{-0.19}$ & \cite{freire2008massive}\\
J1748-2021B & NS-WD & $2.74^{+0.21}_{-0.22}$ & \cite{freire2008massive}\\
J1748-2446I & NS-WD & $1.91^{+0.02}_{-0.10}$ & \cite{kiziltan2013neutron}\\
J1748-2446J & NS-WD & $1.79^{+0.02}_{-0.10}$ & \cite{kiziltan2013neutron}\\
B1802-07 & NS-WD & $1.26^{+0.08}_{-0.17}$ & \cite{thorsett1999neutron}\\
B2303+46 & NS-WD & $1.38^{+0.06}_{-0.10}$ & \cite{thorsett1999neutron}\\
J0740+6620 & NS-WD & $2.14^{+0.10}_{-0.09}$ & \cite{cromartie2020relativistic}\\
J1750-37A & NS-WD & $1.26^{+0.39}_{-0.36}$ & \cite{freire2008massive}\\
J1950+2414 & NS-WD & $1.496^{+0.023}_{-0.023}$ & \cite{zhu2019}\\
J1811-2405 & NS-WD & $2.0^{+0.80}_{-0.50}$ & \cite{Ng2020}\\
J1748-2446am & NS-WD & $1.649^{+0.037}_{-0.11}$ & \cite{andersen2018}\\
J1741+1351 & NS-WD & $1.14^{+0.43}_{-0.25}$ & \cite{arzoumanian2018nanograv}\\
J0045-7319 & NS-MS & $1.58^{+0.34}_{-0.34}$ & \cite{Thorsett1999}\\
J1023+0038 & NS-MS & $1.71^{+0.16}_{-0.16}$ & \cite{deller2012parallax}\\
J1903+0327 & NS-MS & $1.666^{+0.010}_{-0.021}$ & \cite{arzoumanian2018nanograv}\\
\\
\enddata
\tablecomments{95 neutron star mass measurements from \cite{Rocha2021} with $1\sigma$ uncertainties, including system name, type and reference. These are also displayed in Figure \ref{fig:nsdotplot}}
\end{deluxetable*}

\begin{figure}[h!]
 \epsscale{0.85}
\plotone{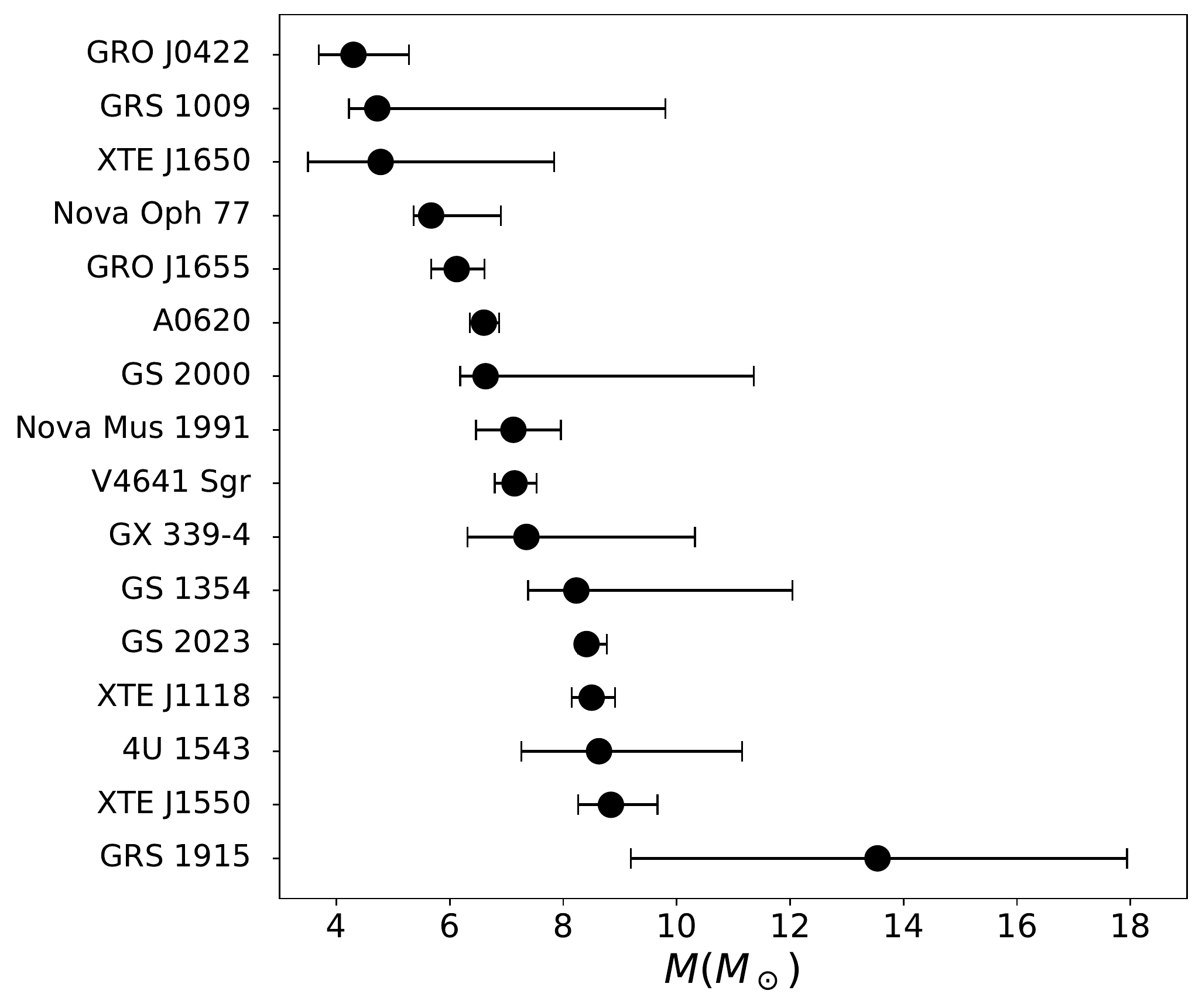}
\caption{Nominal values and $1\sigma$ credibility ranges for the 16 BH mass measurements from \cite{Ozel2010}.\label{fig:bhdotplot}}
\end{figure}

The galactic black hole distribution employed here has been taken from \cite{Ozel2010}, inferred from a sample of 16 BH mass measurements from low-mass X-ray binaries. For a simple Gaussian fit, they obtained mean $\mu=7.8\,\Msun$ and $\sigma=1.2\,\Msun$, from which a lower black hole mass of $\sim 5 M_{\odot}$ is derived. This result is compatible with the simple Gaussian fit by \cite{Farr2011} for 15 of the same low-mass and 5 high-mass X-ray binaries. We list in Table \ref{tab:bhtab} the Gaussian mass distributions corresponding to the 16 measurements in \cite{Ozel2010}, as well as references upon which the distributions were computed. We show in Figure \ref{fig:bhdotplot} the same 16 measurements.

\begin{deluxetable*}{lcl}[h]
\tabletypesize{\scriptsize}
\tablewidth{0pt} 
\tablecaption{Mass Distributions for 16 Black Holes from \cite{Ozel2010}\label{tab:bhtab}}
\tablehead{
\colhead{Name} & \colhead{$M$}& \colhead{Reference} \\
{} & $\left(\Msun\right)$ & {}
} 
\startdata 
GRS 1915 & $AN(13.54, 4.4, 4.35)$ & \cite{Greiner2001} \\
XTE J1550 & $AN(8.84, 0.82, 0.58)$ & \cite{Orosz2011} \\
4U 1453 & $AN(8.63, 2.52, 1.37)$ & \cite{Orosz2003} \\
XTE J1118 & $AN(8.5, 0.41, 0.36)$ & \cite{Gelino2008, Harlaftis2005} \\
GS 2023 & $AN(8.41, 0.36, 0.15)$ & \cite{Charles2006, Khargaria2010} \\
GS 1354 & $AN(8.23, 3.81, 0.85)$ & \cite{Casares2009} \\
GX 339-4 & $AN(7.35, 2.97, 1.04)$ & \cite{Hynes2003,MunozDarias2005} \\
V4641 Sgr & $AN(7.14, 0.39, 0.35)$ & \cite{Orosz2003} \\
Nova Mus 1991 & $AN(7.12, 0.84, 0.66)$ & \cite{Gelino2001} \\
GS 2000 & $AN(6.63, 4.73, 0.45)$ & \cite{Charles2006} \\
A0620 & $AN(6.6, 0.27, 0.25)$ & \cite{Cantrell2010, Neilsen2008} \\
GRO J1655 & $AN(6.12, 0.49, 0.45)$ & \cite{Gelino2001} \\
Nova Oph 77 & $AN(5.67, 1.23, 0.31)$ & \cite{Charles2006} \\
XTE J1650 & $AN(4.78, 3.06, 1.28)$ & \cite{Orosz2004} \\
GRS 1009 & $AN(4.72, 5.08, 0.5)$ & \cite{Filippenko1999} \\
GRO J0422 & $AN(4.3, 0.98, 0.61)$ & \cite{GelinoHarrison2003}
\\
\enddata
\tablecomments{In each line we identify the name of object, the individual mass distribution employed and the source for the data used in computing the mass. $N(\mu,\sigma)$ indicates a normal distribution with mean $\mu$ and standard deviation $\sigma$, while $AN(\mu, \sigma_1, \sigma_2)$ an asymmetrical normal distribution with peak at $\mu$, standard deviation $\sigma_1$ above $\mu$ and standard deviation $\sigma_2$ below. }
\end{deluxetable*}

\clearpage

\bibliography{bibliography}{}
\bibliographystyle{aasjournal}

\end{document}